\documentclass[conference]{IEEEtran}
\IEEEoverridecommandlockouts
\usepackage{hyperref}
\usepackage{amssymb}
\usepackage{amsthm}
\usepackage{multirow}
\usepackage{inputenc}
\usepackage{enumerate} 
\usepackage{textcomp}
\usepackage{listings}
\usepackage{array}
\usepackage{tikz}
\usetikzlibrary{shapes.geometric, arrows}

\usepackage[switch,mathlines]{lineno}
\usepackage{soul}
\usepackage{xfrac}
\usepackage{bbm}
\usepackage{breqn}
\usepackage{graphicx}
\usepackage{cite}
\usepackage{algorithm,algpseudocode}
\algnewcommand{\Inputs}[1]{%
  \State \textbf{Inputs:}
  \Statex \hspace*{\algorithmicindent}\parbox[t]{.8\linewidth}{\raggedright #1}
}
\algnewcommand{\Initialize}[1]{%
  \State \textbf{Initialize:}
  \Statex \hspace*{\algorithmicindent}\parbox[t]{.8\linewidth}{\raggedright #1}
}
\makeatletter 
\newcommand{\linebreakand}{%
  \end{@IEEEauthorhalign}
  \hfill\mbox{}\par
  \mbox{}\hfill\begin{@IEEEauthorhalign}
}
\makeatother 

\usepackage{algpseudocode}

\usepackage{cancel}
\usepackage{cite}
\usepackage{bm}
\usepackage{subfig}
\usepackage{float}
\def\BibTeX{{\rm B\kern-.05em{\sc i\kern-.025em b}\kern-.08em
    T\kern-.1667em\lower.7ex\hbox{E}\kern-.125emX}}

\begin{document}
\title{Quantifying Non-linear Dependencies in Blind Source Separation of Power System Signals using Copula Statistics \\
\thanks{The authors gratefully acknowledge the financial support of NSF via grant ID 1917308.}
}
\author{\IEEEauthorblockN{Pooja Algikar}
\IEEEauthorblockA{\textit{Electrical and Computer Engineering}\\
\textit{Northern Virginia Center, Virginia Tech}\\
apooja19@vt.edu}\\
\and
\IEEEauthorblockN{Lamine Mili}
\IEEEauthorblockA{\textit{Electrical and Computer Engineering} \\
\textit{Northern Virginia Center, Virginia Tech}\\
lmili@vt.edu}\\
\and
\IEEEauthorblockN {Kiran Karra}
\IEEEauthorblockA{\textit{Protocol Labs} \\
\textit{Los Angeles, CA, USA}\\
kiran.karra@gmail.com}\\
\and
\linebreakand 

\IEEEauthorblockN {Akash Algikar}
\IEEEauthorblockA{\textit{Synopsys Inc}\\
\textit{Hyderabad, Telangana, India}\\
algikara@gmail.com}\\
\and

\IEEEauthorblockN {Mohsen Ben Hassine}
\IEEEauthorblockA{\textit{Department of Computer Science}\\
\textit{University of El Manar, Tunisia}\\
mohsenmbh851@gmail.com}\\

}
\maketitle
\begin{abstract}
The dynamics of a power system with a significant presence of renewable energy resources are growing increasingly nonlinear. This nonlinearity is a result of the intermittent nature of these resources and the switching behavior of their power electronic devices. Therefore, it is crucial to address these non-linearity in the blind source separation methods. In this paper, we propose a blind source separation of a linear mixture of dependent sources based on 
copula statistics that measure the non-linear dependence between source component signals structured as copula density functions. The source signals are assumed to be stationary. The method minimizes the Kullback-Leibler divergence between the copula density functions of the estimated sources and of the dependency structure. The proposed method is applied to data obtained from the time-domain analysis of the classical 11-Bus 4-Machine system. Extensive simulation results demonstrate that the proposed method based on copula statistics converges faster and outperforms the state-of-the-art blind source separation method for dependent sources in terms of interference-to-signal ratio.
\end{abstract}

\begin{IEEEkeywords}
Blind source separation, Archimedean copula, dependent sources, under-determined system 
\end{IEEEkeywords}

\section{Introduction}

The process of detecting anomalies and disturbances involves separating the measured signals into independent sources, commonly referred to as blind source separation. This term stems from the fact that we only have access to the observed signals. In the context of moving toward a net-zero energy landscape and integrating renewable energy resources, the relationships between source signals become more complex. These dependencies exhibit higher-order characteristics, which classical blind source separation techniques, designed to account for only second-order dependencies, may not effectively address.

Blind source separation (BSS) algorithms are fundamentally based on the principles of mutual independence, temporal structures, time-frequency analysis, non-stationarity, uncorrelation, and matrix factorization. Temporal analysis methods are effective in the special case where each source signal occurs alone in large time intervals. The BSS problem aims at recovering the underlying source signals from the measured mixture signals via the estimation of the de-mixing matrix.

The prime BSS techniques such as independent component analysis (ICA) and principal component analysis (PCA), to name a few, exploit the mutual independence assumption of source signals. Methods for performing ICA include framing the cost function that measures the non-Gaussianity of the estimated sources based on higher order moments and cumulants, maximization of information transfer, negentropy maximization, and non-linear PCA \cite{Clifford1995ChapterAnalysis}. In second-order blind separation methods \cite{Belouchrani1997AStatistics}, the assumption of independence is relaxed to uncorrelation. In various areas of power system analysis, including  fault detection and localization, load monitoring, anomaly detection, power quality monitoring , and energy management, we often confront the complexities arising from higher-order dependencies between source signals. For example, in modern power grids, as the penetration of renewable generation units increases, we face the challenge of increased non-linearity in the system due to intermittency of renewable generators, switching and harmonic behaviour of inverter-based resources, charging and discharging dynamics of energy storage devices, and complex control strategies for decisions based on real-time data, etc \cite{Shair2021PowerElectronics}.

Several methods have been proposed in the literature accounting for the dependency between the sources including (i) short-term Fourier transform \cite{Abrard2005ASources}, (ii) continuous wavelet transform  \cite{Lee1999BlindRepresentations}, and (iii) minimization of mutual information and/or divergence as a measure of dependence in BSS model \cite{Li2007EfficientSources,Pham2002MutualSources,ElRhabi2004AMixtures,ElRhabi2013ANoise,Candan2005LecturePreface,Ma2007CopulaAnalysis,Keziou2014NewSources}.
Minimization of mutual information methods relax the assumption of independence by assuming a certain dependency structure attributed to the source components modeled as a copula, where semi-parametric likelihood is optimized for its model selection. More recently, the blind source separation algorithms have been developed to solve underterminancy \cite{Lu2023AnSources}, sparseness in the sources\cite{Lu2023AnSources,Kemiha2023Single-ChannelReconstruction}, correlated sources \cite{PehlevanBiologically-PlausibleSources}, latent sources \cite{Wu2022BayesianProcess}. However, limited research has focused to Blind Source Separation (BSS) algorithms that specifically target dependencies by emphasizing the parametric learning of higher-order dependencies in sources. \cite{Erdogan2022ANSOURCES} proposed information maximization based BSS method for dependent and independent signals. 

In the dynamic landscape of power distribution grids, the dominant oscillation modes following a contingency exhibit time-varying characteristics. This transitional behavior involves intricate nonlinear interactions and couplings, stemming from both self-interactions and mutual influences among system modes. As a result, conventional second-order-based BSS techniques may not be directly applicable to address these complexities. Moreover, improved source signal estimates lead to more precise identification of oscillation modes.

The aim of this paper is to develop an algorithm using copula statistics (CoS) \cite{BenHassine2017ALearning} to model the dependency structure of the estimates of source signals regressed upon which is the parameter of dependence. The Copula statistics (CoS) is a bi-variate index that estimates the non-linear dependence between random variables. We employ a Kullback-Leibler (KL) divergence-based semi-parametric optimization solved using the gradient descent algorithm to estimate the dependent source signals. A regression model is formed to estimate the parameter of dependency structure for each type of parametric copula model from the Archimedean family. The proposed algorithm CoS-based Copula Component Analysis (CCCA) can be extended to the case of independent sources by replacing their dependency structure with the copula of independence. We assess the performance of the proposed algorithm on a simple $2 \times 2$ mixing matrix where the source signals are sampled from Gumbel, Clayton, Frank, and Gaussian copula. Additionally, the algorithm is applied to conduct a time-domain analysis of a 4-machine 9-bus system. The specific problem addressed involves the separation of rotor angular velocities as the source signals. We assess the effectiveness of the proposed CCCA by comparing it with copula-based component analysis (CCA) \cite{Ma2007CopulaAnalysis,Keziou2014NewSources} considering the similarity with the method. The evaluation metrics employed for the performance analysis is the Signal-to-Noise Ratio (SNR).
\section{Background}
\subsection{A Brief Introduction to Copula}
A copula is a function that specifies the joint distribution of two or more uniformly distributed random variables. Let us consider a $p$-dimensional random vector $(X_{1},\hdots,X_{p})$ with marginal cumulative distributions $F_{X_{1}}(X_{1}),\hdots,F_{X_{p}}(X_{p})$ respectively. A copula is a function of marginal probability distribution functions. Consequently, it is a joint cumulative probability distribution function that provides a dependency structure between the random variables $(X_{1},\hdots, X_{p})$. Formally, we have a joint cumulative probability distribution function as a function of copula represented as 
$F_{\bm{X}}(\bm{x})  = \bm{C}_{\bm{X}}(F_{X_1}(x_1),F_{X_2}(x_2),\hdots,F_{X_p}(x_p)),$
and copula density of random vectors $X_{1},\hdots,X_{p}$ as 
\begin{equation}
\small
   \bm{c}_{\bm{X}}(F_{X_1}(x_1),\hdots,F_{X_p}(x_p))= \frac{\partial^{p} C_{\bm{X}}(F_{X_1}(x_1),\hdots,F_{X_p}(x_p)}{\partial F_{X_1}(x_1) \cdots F_{X_p}(x_p)}.
\end{equation}
Similarly, the joint probability density function of random vectors, $f_{\bm{X}(\bm{x})}$ with marginal probability density functions, $f_{X_{1}}(x_{1}),\hdots, f_{X_{p}}(x_{p})$ is given by $f_{\bm{X}(\bm{x})}=\prod_{i=1}^{p} f_{X_i}(x_i)c_{\bm{X}}\left(F_{X_1}(x_1),\hdots,F_{X_p}(x_p)\right) $

\subsection{Background on Independent Component Analysis}
Let us consider the independent sources, $\bm{s}_{i}\in \mathbb{R}^q$ and their observations $\mathbf{x}(t)\in \mathbb{R}^p$ at $t=1,\hdots,n$ instances. Let $\mathbf{A}\in \mathbb{R}^{p\times p}$ be an invertible mixing matrix. Let us assume that the observed signals gathered into the matrix,  $\mathbf{X}\in\mathbb{R}^{p\times n}$, are linearly related to each of the source signals as $ \mathbf{X}= \mathbf{W}\mathbf{S},$ where $\mathbf{W}\in\mathbb{R}^{q\times p}$ is the mixing matrix.\\
Corollary 1: 
According to Darmois theorem, if the underlying source components $(\bm{s}_{1}(t),\hdots,\bm{s}_{q}(t));\;\forall t=1,\hdots,n$ are mutually independent and at most one of the components is Gaussian, then a consistent estimator of $\mathbf{W}$ is the one that makes the components of the estimated source vector $y(t)$ statistically independent.
Several versions of ICA are presented in the literature, which considers the dependency structure between the sources \cite{Hyvarinen2017NonlinearSources,Aguilera2013BlindAlgorithm}.

\section{Dependent Component Analysis using Copula}
We formulate the problem of de-mixing matrix estimation using the copula function that models the statistical dependency among the estimates of source vectors,  $\{{y}_{1}(t),\hdots,{y}_{p}(t)\}$. 
When the sources $\{s_{1}(t),\hdots,s_{p}(t)\}$ are dependent, the dependency structure is captured using the prior information available about the sources. We assume that the dependency structure is represented by an unknown semi-parametric copula, $\bm{C}_{\bm{S}}(\cdot;\bm{\alpha})$ where $\bm{\alpha}\in\mathbb{R}^{p}$ is known as the parameter of dependence. The separation method is based on the minimization with respect to $\bm{c}_{\bm{Y}}$ of a non-parametric estimate of the Kullback-Liebler divergence between the probability density functions of the non-parametric copula $\bm{c}_{\bm{Y}}(\cdot)$ of the vector $\bm{Y}=\mathbf{W}\mathbf{X}$ and semi-parametric copula density $\bm{c}_{\bm{S}}(\cdot,\bm{\alpha})$ of the vector of sources $\bm{S}$. Formally, we have 
\begin{dmath}
\small
    \textrm{KL}(\bm{c}_{\bm{Y}}||\bm{c}_{\bm{S}};\bm{\alpha})=\int_{[0\; 1]^{p}}\textrm{log}\left(\frac{\bm{c}_{\bm{Y}}(F_{Y_{1}}(y_{1}),\hdots,F_{Y_{p}}(y_{p}))}{\bm{c}_{\bm{S}}(\cdot,\bm{\alpha})} \right)\\
    \bm{c}_{\bm{Y}}(F_{Y_{1}}(y_{1}),\hdots,F_{Y_{p}}(y_{p}))d(F_{Y_{1}}(y_{1})\hdots F_{Y_{p}}(y_{p}))
    \end{dmath}
    \begin{align}    &=\mathbb{E}\left[\textrm{log}\left(\frac{\bm{c}_{\bm{Y}}(F_{Y_{1}}(y_{1}),\hdots,F_{Y_{p}}(y_{p}))}{\bm{c}_{\bm{S}}(F_{Y_{1}}(y_{1}),\hdots,F_{Y_{p}}(y_{p}),\bm{\alpha})}\right)\right].
\end{align}
To achieve this separation, the idea is to minimize a statistical estimate of $\widehat{\textrm{KL}}(\bm{c}_{\bm{Y}}||\bm{c}_{\bm{S}};\bm{\alpha})$, with respect to $\mathbf{W}$, constructed from the data $(\bm{y}_{1},\hdots,\bm{y}_{T})$. The de-mixing matrix is then estimated by 
\begin{equation}\label{1}
\small
    \widehat{\mathbf{W}}=\underset{\mathbf{W}}{\mathrm{arg\;min}}\; \widehat{\textrm{KL}}(\bm{c}_{\bm{Y}}||\bm{c}_{\bm{S}};\bm{\alpha}). 
\end{equation}
The estimate of the criteria $\widehat{\textrm{KL}}(\bm{c}_{\bm{Y}}||\bm{c}_{\bm{S}};\bm{\alpha})$ is obtained by using
\begin{equation}
\small
    \widehat{\textrm{KL}}(\bm{c}_{\bm{Y}}||\bm{c}_{\bm{S}};\bm{\alpha})=\frac{1}{T}\sum_{i=1}^{T}\textrm{log}\left(\frac{\bm{c}_{\bm{Y}}(\widehat{F}_{Y_{1}}(y_{1}(i)),\hdots,\widehat{F}_{Y_{p}}(y_{p}(i)))}{\bm{c}_{\bm{S}}(\widehat{F}_{Y_{1}}(y_{1}(i)),\hdots,\widehat{F}_{Y_{p}}(y_{p}(i)),\bm{\alpha})} \right),
\end{equation}
where
\begin{equation}\label{2}
\small
    \widehat{\bm{c}}_{\bm{Y}}(\bm{u})=\frac{1}{TH_{1}\cdots H_{p}}\sum_{n=1}^{T}\prod_{i=1}^{p}k\frac{\widehat{F}_{Y_{i}}(Y_{i}(n))-u_{i}}{H_{i}},\; \forall\bm{u}\in [0\; 1]^{p},
\end{equation}
is the kernel estimator of the copula density $\bm{c}_{\bm{Y}}(\cdot)$, and $\widehat{F}_{Y_{i}}(z),i=1,\hdots,p,$ is the estimate of the marginal distribution function, $F_{Y_{i}}(z)$, of the random variable $Y_{i}$ in point $z\in\mathbb{R}$, defined by
\begin{equation}\label{3}
    \widehat{F}_{Y_{i}}(z)=\frac{1}{T}\sum_{n=1}^{T}K\left(\frac{Y_{i}(n)-z}{h_{i}}\right),\; \forall i=1,\hdots,p,
\end{equation}
where $K(\cdot)$ is the primitive of the kernel $k(\cdot)$. The kernel function we chose is the standard Gaussian density with zero expectation and unit variance, where the smoothing windows $h_{i}$ and $H_{i}$ are given as 
   $ h_{i}=\frac{4}{3}^{0.2}T^{\frac{-1}{5}}\widehat{\sigma_{i}},
     H_{i}=\left(\frac{4}{p+2}\right)^{\frac{1}{p+4}}T^{\frac{-1}{p+4}}\widehat{\Sigma_{i}},$ respectively.
Here, $\widehat{\sigma_{i}}$ denotes the  empirical standard deviation of the sample $(Y_{i}(1),\hdots,Y_{i}(T))$ whereas  $\widehat{\Sigma_{i}}$ denotes the empirical standard deviation of $(\widehat{F}_{Y_{i}}(1),\hdots,\widehat{F}_{Y_{i}}(T));\; i=1,\hdots,p$ \cite{GhazdaliBlindCopula}.  
\section{Method}
We propose a method to estimate the parameter of dependency $\bm{\alpha}$ prior to the estimation of de-mixing matrix $\mathbf{W}$ that models the dependence structure of the source signals in terms of the given observed signals $\mathbf{X}$ using the statistical copula index CoS. We then propose a modified KL-divergence algorithm to estimate the de-mixing matrix for the source separation in which the statistical structure of dependence as well as independence of the source signals is modelled using the CoS.  

\subsection{Estimation of Dependency Parameter of Copula Structure}
The copula statistics, CoS, is a statistical index that measures the strength of bi-variate dependence including linear, cubic, and fourth root dependence. The corresponding measure relies on properties such as concordance, quadrant dependence, and comonotonicity introduced by Lehmann [25] between two random variables in terms of a relative distance function between the empirical copula, the Fréchet-Hoeffding bounds and the independence copula. The CoS approximates the Pearson’s correlation coefficient for the Gaussian copula and the Spearman’s correlation coefficient for many copulas for large sample sizes. It ranges from zero to unity, attaining its lower and upper limit for the cased of independence and functional dependence, respectively. The CoS estimation algorithm is given in Algorithm \ref{alg1}.
\begin{algorithm}[!htbp]
\small
\caption{Algorithm for Estimating CoS between two random variables $X$ and $Y$}
\begin{algorithmic} [1]
\State Calculate $u_{j}$, $v_{j}$ and $C_{n}(u,v)$ as follows: 
\begin{itemize}
    \item $u_{j}=\frac{1}{n}\sum_{j=1}^{m}\mathbbm{1}(k\neq j:x_{k}\leq x_{j})$
    \item $v_{j}=\frac{1}{n}\sum_{j=1}^{m}\mathbbm{1}(k\neq j:y_{k}\leq y_{j})$
    \item $C_{n}(u,v)=\frac{1}{n}\sum_{j=1}^{m}\mathbbm{1}(u_{j}\leq u, v_{j}\leq v)$
\end{itemize}
\State Order $x_{j}^{'}$s to get $x(1)\leq \hdots \leq x(n)$, which results in  $u(1)\leq \hdots \leq u(n)$ since $u_{j}=\frac{R_{x_{j}}}{n}$, where $R_{x_{j}}$ is the rank of $x_{j}$;
\State Determine the domains $\mathcal{D}_{i};i=1,\hdots,m,$ where each $\mathcal{D}_{i}$ is a u-interval associated with a non-decreasing or non-increasing sequence of $C_{n}(u_{j},v_{p}), j=1,\hdots,n$; 
\State Determine the smallest and the largest value of $C_{n}(u,v)$, denoted by $C_{i}^{min}$ and $C_{i}^{max}$, and find the associated $u_{i}^{min}$ and $u_{i}^{max}$ for each domain $\mathcal{D}_{i};i=1,\hdots,m,$;
\State Calculate $\lambda(C_{i}^{min})$ and $\lambda(C_{i}^{max})$ as 
\begin{equation*}
\small
    \lambda(C(u,v))=\begin{cases}
    \frac{C(u,v)-uv}{Min(u,v)-uv},&  \textrm{if}\; C(u,v)\geq uv,\\
    \frac{C(u,v)-uv}{Max(u+v-1,0)-uv}, & \textrm{if}\; C(u,v)< uv;
\end{cases}
\end{equation*}
\State If $\lambda(C_{i}^{min})$ and $\lambda(C_{i}^{max})$ are equal to one, go to step 8; 
\State Calculate the absolute difference between the three consecutive values of $C_{n}(u_{(i)},v_{j})$ centered at $u_{i}^{min}$ (respectively at $u_{i}^{max}$) and decide that the central point is a local optimum if (i) both absolute differences are smaller than or equal to $\frac{1}{n}$ and (ii) there are more than four points within the two adjacent domains, $\mathcal{D}_{i}$ and $\mathcal{D}_{i+1}$;
\State Calculate $\gamma_{i}$ given by 
\begin{equation*}
\small
   \gamma_{i}=\begin{cases}
    1,&  \textrm{at a local optimum of}\; Y=f(X) \textrm{on}\; D_{i},\\
    \frac{\lambda(C_{i}^{min})+\lambda(C_{i}^{max})}{2}, & \textrm{otherwise;}
\end{cases}.
\end{equation*}
\State Repeat Steps 2 through 7 for all the m domains, $\mathcal{D}_{i};i=1,\hdots,m,$;
\State Calculate the CoS given by 
    $CoS(X,Y)=\frac{1}{n+m-1}\sum_{i=1}^{m}n_{i}\gamma_{i}.$ 
\end{algorithmic}\label{alg1}
\end{algorithm}

The aim is to estimate the parameter of dependence of the copula density function, ${\alpha}$, using CoS. The bi-variate nature of the relationship between the CoS and the parameter of interest $\alpha$ is studied for the family of Archimedian copula. For Gaussian copula, it was observed that the CoS is approximately equal to the Pearson's correlation coefficient. For the other copula, it is observed that $\alpha$ exhibits a quadratic relationship with CoS. Therefore, we develop a linear regression model as follows:
\begin{equation}\label{20}
 \alpha=a_{1}\textrm{CoS}^{2}+a_{2}\textrm{CoS}+a_{3}.
\end{equation}
A least squares estimator is applied to estimate  the coefficients of the regression model $\widehat{\bm{a}}=\underset{{\bm{a}}}{\mathrm{arg \;min}}\; ||\bm{\alpha}-\mathbf{H}^{T}\bm{a}|| $ with a quadratic basis function $\mathbf{H}=[\textrm{CoS}_{i}^{2}, \textrm{CoS}_{i}, 1]^{T}; \; i=1,\hdots,T.$ The training data set consists of CoS estimated at $5000$ samples of two random variables, which are generated with $\alpha$ ranging from -0.99 to 0.99 for the Gaussian, 0.001 to 20 for Frank and Clayton copula, and from 1 to 20 for the Gumbel copula.
\subsection{Algorithm for the Source Separation }
Let us consider the case where there is a priori information about the source dependence structure, represented by a semi-parametric copula model $C_{\bm{S}}(\cdot;\bm{\alpha})$ with an unknown parameter of dependence $\bm{\alpha}$. This scenario is common in many practical applications.
Let $\widehat{{\theta}}_{ij}=\textrm{CoS}(\mathbf{x}_{i},\mathbf{x}_{j});\; \textrm{for}\; i\neq j;i=1,\hdots,p;j=1,\hdots,p,$ be the estimated copula statistics between the observations $\{\mathbf{x}_{i},\mathbf{x}_{j}\}$ using the copula statistics estimation procedure. We then estimate the dependence parameter, $\alpha_{ij}$, using the previously estimated coefficients of regression $\bm{a}$ as $\widehat{{\alpha}_{ij}}=\mathbf{h}\widehat{\bm{a}}$, where  $\mathbf{h}=[\widehat{\theta_{ij}}^{2}, \widehat{\theta}_{ij}, 1]^{T}$ forms the quadratic basis function of ${\theta_{ij}}$. Because  $\alpha_{ij}=\alpha_{ji}$, the final estimate of $\bm{\alpha}, \widehat{\bm{\alpha}}\in \mathbb{R}^{p}$ measures the dependence parameter between the set of random observations $\{\mathbf{x}_{1},\hdots,\mathbf{x}_{p}\}$ mutually.
Once the parameter for dependence $\bm{\alpha}$ is estimated, a gradient descent algorithm for source separation is employed. The modified estimate of KL divergence is given by 
$\widehat{\textrm{KL}}(\bm{c}_{\bm{Y}}||\bm{c}_{\bm{S}};\bm{\alpha},\bm{\theta})=\frac{1}{T}\sum_{i=1}^{T}\textrm{log}\left(\frac{\bm{c}_{\bm{Y}}(\widehat{F}_{Y_{1}}(y_{1}(i)),\hdots,\widehat{F}_{Y_{p}}(y_{p}(i)))}{\bm{c}_{\bm{S}}(\widehat{F}_{Y_{1}}(y_{1}(i)),\hdots,\widehat{F}_{Y_{p}}(y_{p}(i)),\bm{\alpha};\bm{\theta})} \right),$
and the de-mixing matrix is estimated as
\begin{equation}\label{4}
\small
    \widehat{\mathbf{W}}=\underset{\mathbf{W}}{\mathrm{arg\;min}}\; \widehat{\textrm{KL}}(\bm{c}_{\bm{Y}}||\bm{c}_{\bm{S}};\bm{\alpha},\bm{\theta}). 
\end{equation}
The solution for \eqref{4} can be calculated using a gradient descent algorithm given in Algorithm \ref{tab0} with the previously given estimates of copula density in \eqref{2}, and the marginal distribution function of the random variables as shown in \eqref{3}. Since the index CoS is based on copula that attains its lower bound of zero for the case of independent random variables, a common algorithm for source separation for both the cases of dependent and independent sources is proposed. 
\begin{figure}[t]%
    \centering
  {{\includegraphics[height=3cm,width=5cm]{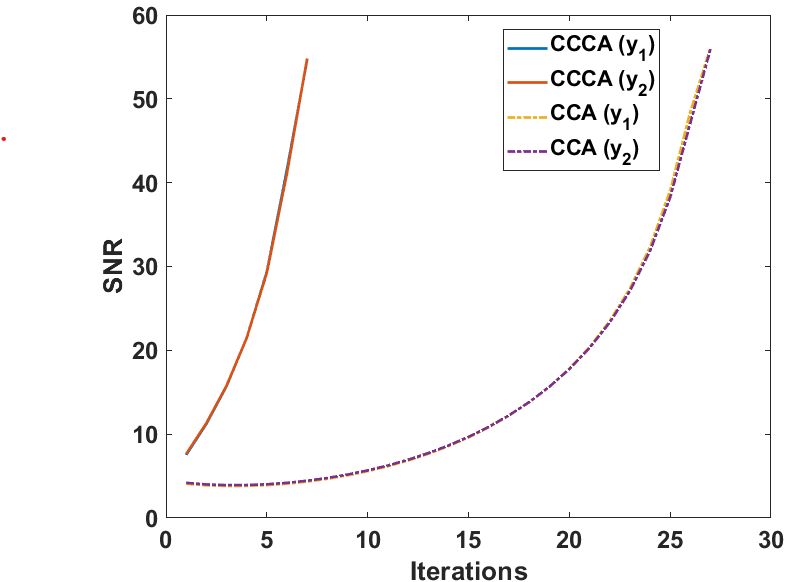}}}
  \caption{The comparison of results obtained from CCCA and CCA for the estimated source signals $s_{1},s_{2}$ denoted as $\mathbf{y}_{1},\mathbf{y}_{2}$, respectively, derived from Gumbel copula family with $\alpha=5$}
    \label{num}
\end{figure}
\begin{algorithm}[!htbp]
\small
\caption{Algorithm for Estimating the De-mixing Matrix $\bm{W}$ }
\begin{algorithmic} [1]
\State Input: $\mathbf{X}$ observation vector, copula density model: $\{\bm{c}_{\bm{\alpha}}(\cdot);\;\alpha\in\mathbb{R}^{p-1}\}$ 
\State Output: $\bm{Y}$ estimated vector of sources
\State Begin: 
\begin{itemize}
    \item Initialization: $\mathbf{W}^{0}=\mathbf{I}_{p},\; \bm{Y}^{0}=\mathbf{W}^{0}\mathbf{X}.$ For estimated $\bm{\alpha}$, $\epsilon>0,\mu>0$.
    \item Loop
    \begin{itemize}
    \item $\bm{Y}^{(k)}=\mathbf{W}^{(k)}\mathbf{X}$
     \item $\widehat{\theta}_{ij}=\textrm{CoS}(\mathbf{y}_{i}^{(k)},\mathbf{y}_{j}^{(k)});\; i,j=1,\hdots,p.$
    \item $\widehat{\alpha}=\mathbf{h}\widehat{\mathbf{a}};\;$ where $\mathbf{h}=[\widehat{\theta}^{2}_{ij} \widehat{\theta}_{ij} 1]$
    \item $\mathbf{W}^{(k+1)}=\mathbf{W}^{(k)}-\mu\frac{\bm{d}\widehat{\textrm{KL}}\left(\bm{c}_{\bm{Y}^{(k)}}||\bm{c}_{\bm{S}};\bm{\alpha},{\bm{\theta}}\right)}{\mathbf{dW}}$
    \item $\bm{Y}^{(k+1)}=\mathbf{W}^{(k+1)}\mathbf{X}$
     \end{itemize}
    \item Until $||\mathbf{W}^{(k+1)}-\mathbf{W}^{(k)}||<\epsilon$
\end{itemize}
\State End 
\State $\bm{Y}=\bm{Y}^{(k+1)}$.
\end{algorithmic}\label{tab0}
\end{algorithm}
\section{Simulation Results}
\subsection{A Numerical Model}
The $T=500$ realizations of the two source signals, $s_1,s_2$, are generated using bi-variate copula density associated with each of the Archimedian family namely, the Gumbel, Clayton, Frank, and Gaussian copula with the parameter ${\alpha}$ as 5, 5, 5,0.7, respectively.
In our experiment, $\mathbf{A}$ is 
$\begin{bmatrix}
    1 & 0.8\\
    0.8 & 1\\
    \end{bmatrix}$.
Both the proposed CCCA and the comparable standard copula component analysis (CCA) are used to recover the components from their mixtures. Without the attention to study the copula model selection, the respective copula family is adopted for both the methods.
Figure \ref{num} presents the outcomes for the scenario involving sources derived from the Gumbel copula.  
We notice that the proposed CCCA achieves a faster convergence rate to attain the same SNR as that of the comparable CCA with reduced computing time. 
\subsection{11-Bus 4-Machine Test System}
\begin{figure*}%
    \centering
  \subfloat[\centering ]{{\includegraphics[height=2.5cm,width=4cm]{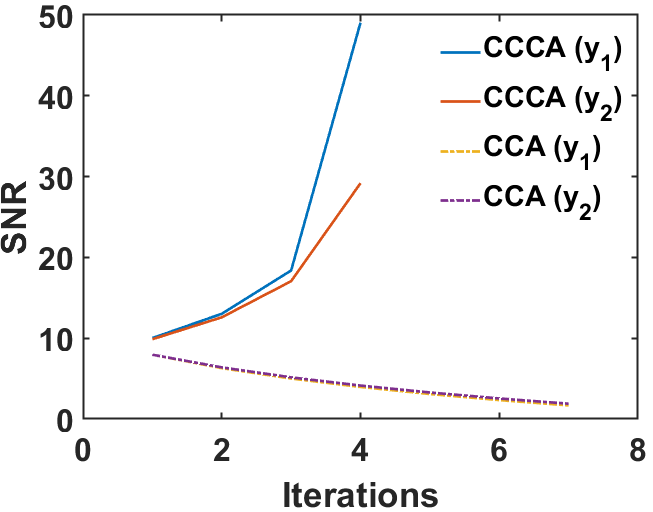}}}%
    \qquad
    \subfloat[\centering ]{{\includegraphics[height=2.5cm,width=4cm]{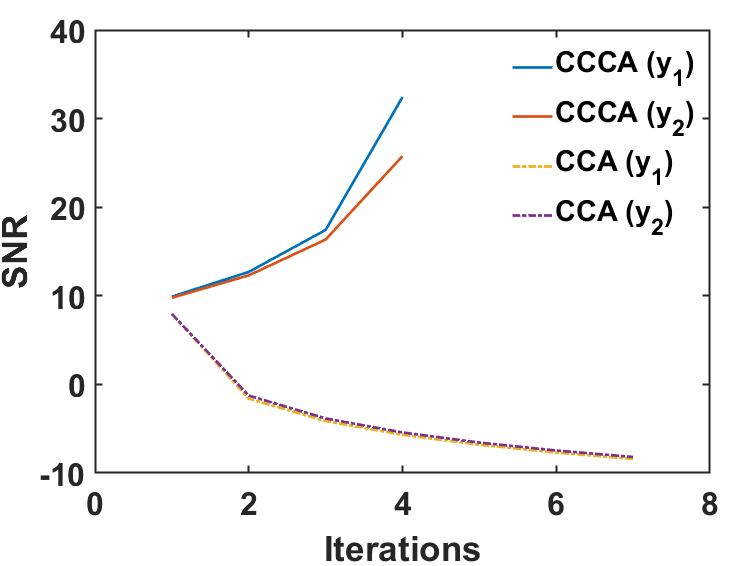}}}%
    \qquad
  \subfloat[\centering ]{{\includegraphics[height=2.5cm,width=4cm]{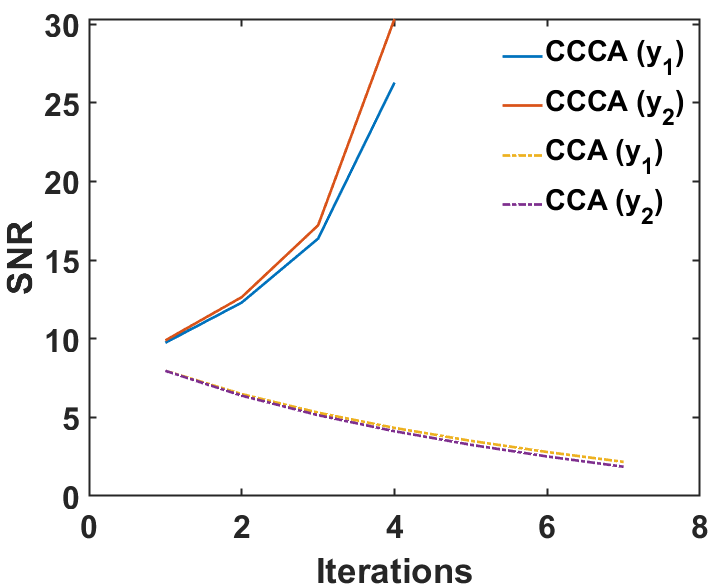}}}%
    \qquad
    \subfloat[\centering ]{{\includegraphics[height=2.5cm,width=4cm]{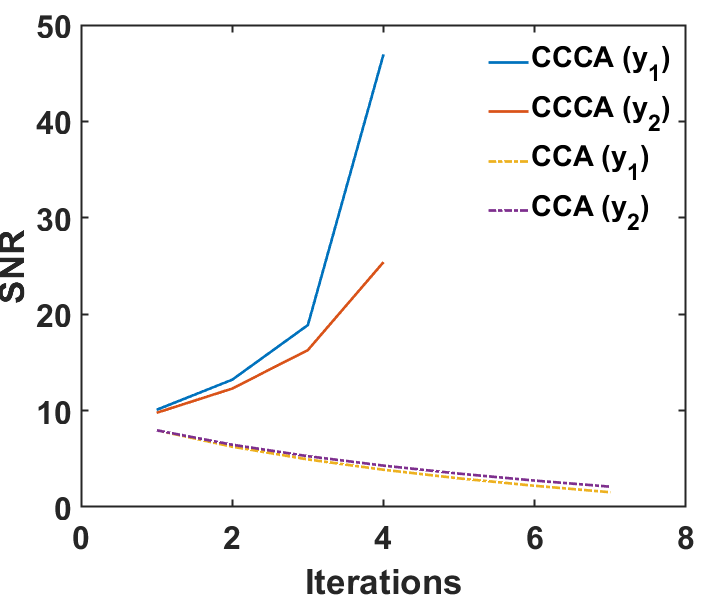}}}%
  \qquad
  \subfloat[\centering ]{{\includegraphics[height=2.5cm,width=4cm]{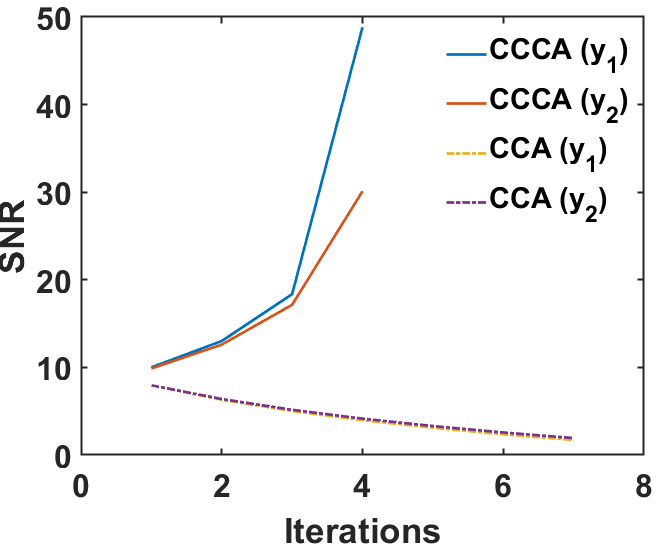}}}%
    \qquad
    \subfloat[\centering ]{{\includegraphics[height=2.5cm,width=4cm]{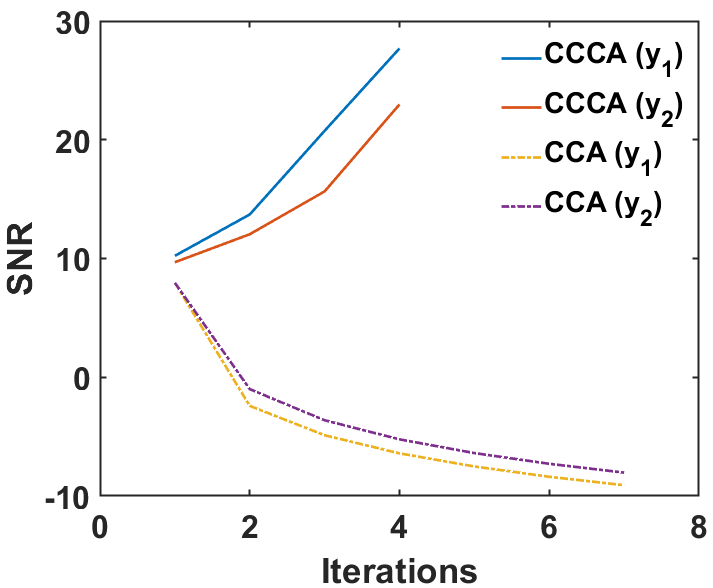}}}%
    \qquad
  \subfloat[\centering ]{{\includegraphics[height=2.5cm,width=4cm]{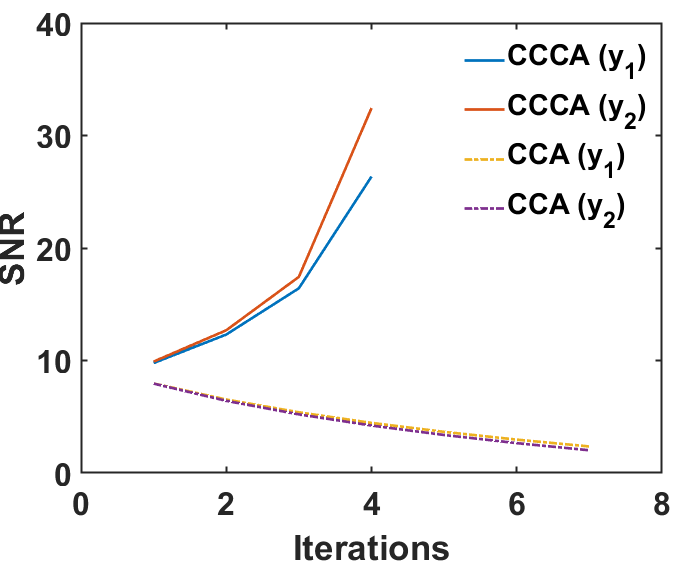}}}%
    \qquad
    \subfloat[\centering ]{{\includegraphics[height=2.5cm,width=4cm]{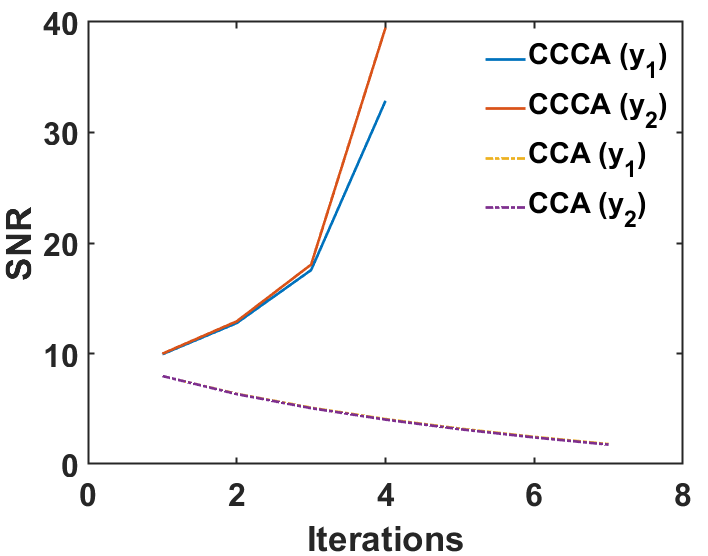}}}%
    \caption{Comparison of the SNR of estimated source signals: rotor angular velocities $\widehat{\omega_{1}}$ and $\widehat{\omega_{2}}$ denoted as $\mathbf{y}_{1},\mathbf{y}_{2}$, respectively, obtained from the proposed CCCA and the CCA for (a) Frank copula fit; (b) Clayton copula fit; (c) Gumbel copula fit; (d) Gaussian copula fit and
    rotor angular velocities $\widehat{\omega_{3}}$ and $\widehat{\omega_{4}}$ denoted as $\mathbf{y}_{1},\mathbf{y}_{2}$ respectively, obtained from the proposed CCCA and the CCA for (e) Frank copula fit; (f) Clayton copula fit; (g) Gumbel copula fit; and (h) Gaussian copula fit.}
    \label{w_34}
\end{figure*}

We consider the simulation case of a time domain analysis with a three-phase unbalanced fault applied on Bus 8 at 1s and cleared at 1.3s. Because of the delay in the clearance, the non-linear coupling between states is dominant in the system. The simulations are performed in Matlab toolbox PSAT carried out on the classical $11$-bus $4$-machine test system \cite{kundur2022power}. The specified sampling frequency and the base frequency are $20$ Hz and  $60$ Hz, respectively. The source signals of interest are rotor angular velocities of the $4$ synchronous generators.
The measured signals are constituted from the mixing matrix $\mathbf{A} = \begin{bmatrix}
    1 & 0.4\\
    0.4 & 1\\
    \end{bmatrix}$ with instantaneously added noise term $n\sim\mathcal{N}(0,0.001)$. 
The demixing matrix $\mathbf{W}$ is initialized to $\mathbf{I}\in \mathbb{R}^{p\times p}$.
In Figure \ref{w_34}, the SNR obtained from CCCA and CCA with the assumption of Frank, Clayton, Gumbel, and Gaussian copula as the dependency structure of the sources is compared. 
It is evident that the CCA fails to converge in the case of strong coupling between the states of interest $\omega_{1},\omega_{2},\omega_{3}, \textrm{and}\; \omega_{4}$ in the case of delayed fault clearance. The proposed CCCA demonstrates a higher degree of generalization compared to CCA. This is substantiated by the fact that CCCA consistently produces superior SNR values. 
The dataset used to obtain the experimental results and code is provided at \href{https://github.com/apooja1/CCCA}{CCCA}. 


\section{Conclusion and Future Work}
Although the simulations are carried out for the case of angular velocities as dependent source signals, the proposed method can be easily extrapolated to the case of independent source signals using the copula of independence. Our simulation results show that the proposed CCCA emerges as a more robust and effective approach for blind source separation problem, particularly when dealing with intricate coupling relationships among the source states. This method finds particular relevance when dealing with complex coupling relationships among the source states, especially in applications like load consumption separation from various sets of consumers. As for the future work, we plan to devise an advanced regression model for the $\alpha$ regression in \eqref{20}.
Furthermore, our future research trajectory involves the development of multivariate CoS index to extend the application to multi-source load separation case to make it scalable for the large-scale power systems. 
\bibliography{references} 
\bibliographystyle{ieeetr}

\end{document}